\documentclass[twocolumn, aps, prl]{revtex4-1}
\usepackage[utf8]{inputenc}
\usepackage{amssymb,amsmath,graphicx,latexsym}
\usepackage{array}
\usepackage{bm}
\usepackage{color}
\usepackage{dcolumn}
\usepackage{multirow}
\usepackage{float}
\usepackage{tabularx}

\usepackage[normalem]{ulem}
\usepackage{hyperref}
\usepackage{grffile}
\usepackage{extarrows}
\usepackage{tikz}
\usepackage{rotating}
\usepackage{pdfpages}
\usepackage[low-sup]{subdepth}

\makeatletter
\AtBeginDocument{\let\LS@rot\@underfined}
\makeatother
\newcommand{\dtabsize}{\footnotesize}

\def\di{\text{d}}

\newcommand{\bea}{\begin{eqnarray}}
\newcommand{\eea}{\end{eqnarray}}
\newcommand{\bp}{\begin{pmatrix}}
\newcommand{\ep}{\end{pmatrix}}
\newcommand{\beq}{\begin{equation}}
\newcommand{\eeq}{\end{equation}}
\newcommand{\bq}{\begin{equation}}
\newcommand{\eq}{\end{equation}}
\newcommand{\ba}{\begin{array}}
\newcommand{\ea}{\end{array}}
\newcommand{\beqa}{\begin{eqnarray}}
\newcommand{\eeqa}{\end{eqnarray}}
\newcommand{\beqs}{\begin{subequations}}
\newcommand{\eeqs}{\end{subequations}}

\def\({\left(}
\def\){\right)}
\def\LB{\left[}
\def\RB{\right]}

\def\leqq{\leqslant}

\def\End{\end{document}}

\begin{document}
\bibliographystyle{apsrev}
\title{Reconciling Hubble Constant Discrepancy from Holographic Dark Energy}

\author{{\sc Wei-Ming Dai}\,$^{1,2}$}
\email{DaiW@ukzn.ac.za}

\author{{\sc Yin-Zhe Ma}\,$^{1,2,3}$}
\email{Ma@ukzn.ac.za}

\author{{\sc Hong-Jian He}\,$^{4,5,6}$}
\email{hjhe@sjtu.edu.cn}

\affiliation{%
$^{1}$School of Chemistry and Physics, University of KwaZulu-Natal,
Westville Campus, Private Bag X54001, Durban, 4000, South Africa\\
$^{2}$NAOC-UKZN Computational Astrophysics Centre (NUCAC),
University of KwaZulu-Natal, Durban, 4000, South Africa\\
$^{3}$Purple Mountain Observatory, CAS, No.10 Yuanhua Road, Qixia District, 
Nanjing 210034, China \\
$^{4}$Tsung-Dao Lee Institute $\&$ School of Physics and Astronomy,\\
Shanghai Key Laboratory for Particle Physics and Cosmology,\\
Shanghai Jiao Tong University, Shanghai 200240, China\\
$^{5}$Institute of Modern Physics $\&$ Physics Department,
Tsinghua University, Beijing 100084, China\\
$^{6}$Center for High Energy Physics, Peking University, Beijing 100871, China
}

\begin{abstract}
\noindent
Holographic dark energy (HDE) describes the vacuum energy in
a cosmic IR region whose total energy saturates the limit of
avoiding the collapse into a black hole.
HDE predicts that the dark energy equation of the state
transiting from greater than the $-1$ regime to less than $-1$, accelerating the
Universe slower at the early stage and faster at the late stage. We propose
the HDE as a new {\it physical} resolution to the Hubble constant discrepancy
between the cosmic microwave background (CMB) and local measurements.
With {\it Planck} CMB and galaxy baryon acoustic oscillation (BAO) data,
we fit the HDE prediction of the Hubble constant as
$H_0^{}\!=\, 71.54\pm1.78\,\mathrm{km\,s^{-1} Mpc^{-1}}$, consistent with local
$H_0^{}$ measurements by LMC Cepheid Standards\,(R19) at $1.4\sigma$ level.
Combining {\it Planck}+BAO+R19, we find the HDE parameter $c=0.51\pm0.02$ and
$H_0^{}\! = 73.12\pm 1.14\,\mathrm{km \,s^{-1} Mpc^{-1}}$,
which fits cosmological data at all redshifts. Future CMB and large-scale structure surveys will further test the holographic scenario.
\\[3mm]
Phys.\ Rev.\ D\,102 (2020) 121302 (Rapid Communication),  
[arXiv:2003.03602 [astro-ph.CO]].
\end{abstract}

\maketitle


\noindent
{\bf\large 1.\,Introduction}
\vspace*{2.5mm}

The cosmological observations
derived from the ``Early'' and ``Late'' Universe
tend to prefer different values of the Hubble constant ($H_{0}$),
leading to a discrepancy between the two types of measurements.
The measurements of the cosmic microwave background (CMB) from {\it Planck} satellite\,\cite{Aghanim:2018eyx} and
Atacama Cosmology Telescope (ACT)\,\cite{2020arXiv200707288A}
measured the Hubble constant to be
$H_{0} = 67.4 \pm 0.5 \mathrm{\,km \,s^{-1} Mpc^{-1}}$ ({\it Planck}), and
$H_0= 67.6 \pm 1.1 \mathrm{\,km \,s^{-1} Mpc^{-1}}$ (ACT).
In contrast, several local measurements give consistent higher values. SH0ES measurement of
Cepheids data (R19) obtained from {\it Hubble Space Telescope} (HST)
gives the improved local determination
$H_0^{}\!= 74.03 \pm 1.42 \mathrm{\,km \,s^{-1} Mpc^{-1}}$~\cite{Riess:2019cxk}.
Replacing the Cepheids with the oxygen-rich Miras discovered in NGC4258, Ref.\,\cite{Huang:2019yhh} measured the Hubble constant
$\,H_0^{}\! = 73.3 \pm 3.9\,\mathrm{km \,s^{-1} Mpc^{-1}}$.
Using the geometric distance to the megamaser-hosting galaxies CGCG 074-064 and NGC 4258, Ref.\,\cite{Pesce20} gives
$H_{0}^{}\!=73.9 \pm 3.0\,{\rm km}\,{\rm s}^{-1}{\rm Mpc}^{-1}$.
In a complementary probe of using gravitationally lensed quasars
with measured time delays in a flat $\Lambda\mathrm{CDM}$ cosmology,
the H0LiCOW team found
$H_{0}= 73.3^{+1.7}_{-1.8}\,\mathrm{km \,s^{-1} Mpc^{-1}}$~\cite{Wong:2019kwg}
and more recently $82.4^{+8.4}_{-8.3}$~\cite{Jee20}.
A combination of different local measurements yields
$H_0^{}\!=73.1\pm 0.9\,\mathrm{km \,s^{-1} Mpc^{-1}}$,
which is $5.7\sigma$ discrepant from the aforementioned
{\it Planck} CMB result\,\cite{Verde:2019ivm}.

\vspace*{1mm}

Various theories have been proposed to resolve this discrepancy,
mainly from two prospects\,\cite{Knox:2019rjx}:
(i)~modifying the early-universe physics to shrink down the sound horizon
at the drag epoch $r^{\rm drag}_{\rm s}$~\cite{Aylor19},
such as including neutrino self-interactions to delay
its free-streaming\,\cite{Kreisch19,He:2020zns}\footnote{%
However, a recent paper showed that this mechanism cannot fully resolve
Hubble tension~\cite{Jedamzik20}.}.
(ii)~modifying dark energy (DE) evolution, by considering the dark sector
interactions\,\cite{DiValentino:2017iww}
or early dark energy (EDE) component\,\cite{Poulin19}.
But these new models are either phenomenologically contrived, or hard to falsify because they always recover $\Lambda$CDM behaviour if some parameters are tuned. For example, the interacting dark sector model assumes a somewhat arbitrary form of interaction, with a free coupling. If the constraints are improved, the interaction coupling is asymptotically approaching zero but can always have a small value, making it almost impossible to rule out the model. The EDE model inserts a short period of fast expansion around recombination,
by demanding ``just about'' the amount of dark energy
to expand the Universe and exit at the right time.
So the EDE scenario, by its construction, is highly contrived.


In this work, we propose the holographic dark energy (HDE) as a striking new resolution to the $H_0^{}$ discrepancy. This scenario is built upon
the holographic principle of quantum gravity from 't~Hooft\,\cite{tHooft:1993dmi} and Susskind\,\cite{Susskind:1994vu},
and the Bekenstein-Hawking entropy bound\,\cite{PhysRevD.7.2333,hawking1975}
that connects an ultraviolet (UV) scale in quantum states to an infrared (IR) cutoff in the
macroscopic scale\,\cite{Cohen99}.
We demonstrate that with a single HDE parameter $\,c\simeq 0.5$\,,
its equation of state (EoS) transits from $\,w>-1$\, to $\,w<-1$,\,
which can naturally resolve the $H_0^{}$ discrepancy between the CMB and local measurements.
We also show that for any value of $c$,
the HDE EoS always evolves as a function of time and never mimics
Einstein's cosmological constant $\Lambda$,
which can be substantiated or falsified with future data.

\vspace*{4mm}

\noindent
{\bf\large 2.\,Holographic Dark Energy}
\vspace*{2.5mm}

In the black hole thermodynamics,
the Bekenstein entropy bound states
that the maximum entropy in a box of volume $L^{3}$
grows only as the box's area\,\cite{PhysRevD.7.2333,hawking1975}.
Then, 't Hooft\,\cite{tHooft:1993dmi} and Susskind\,\cite{Susskind:1994vu} observed
that due to the Bekenstein bound the 3+1 dimensional field theories over-count the
degrees of freedom (d.o.f), which are proportional to the area of
the box surface. They conjectured the holographic principle:
the total d.o.f of any effective field theory
in a box of size $L$ must be below the black hole entropy of the same size,  $L^{3}\Lambda^{3}\!\leqq\! S_{\rm BH}\!=\!\pi L^{2}M^{2}_{\rm Pl}$\,,
where $M_{\rm Pl}^{}\!=(8\pi G)^{-1/2}$ is the reduced Planck mass
and $\Lambda$ denotes the UV cutoff.
Subsequently, Cohen, Kaplan and Nelson\,\cite{Cohen99}
found that if this condition holds,
it implies that there are many states within the box size $L$
with Schwarzschild radii even {\it larger} than $L$.
Hence these states should have collapsed and cannot exist.
To avoid such catastrophe, Ref.\,\cite{Cohen99} tightened up the bound
by imposing a constraint on the IR cutoff $1/L$
that excludes all states lying within the Schwarzschild radius,
$L^{3}\Lambda^{4} \!\leqq\! LM^{2}_{\rm Pl}$\,.
This condition predicts the maximum entropy
$\,S_{\rm max}^{} \!=\! S_{\rm BH}^{3/4}$\,
and is indeed a tighter bound.
In sum, this new holographic condition means
that for quantum states in a box size $L$ to exist without collapsing,
the short-distance UV scale is connected to a long-distance IR cutoff
due to the limit set by the black hole formation.
Namely, the maximum total energy set by the UV cutoff
$\Lambda$ in a region of size $L$ should not exceed the mass of
a black hole with the same size,
$L^{3}\rho_{\Lambda}^{} \!\lesssim\! LM^{2}_\text{Pl}$\,.\,
Hence, the dark energy density is bounded by
$\,\rho_{\Lambda}^{} \!\lesssim M^2_\text{Pl}L^{-2}$.
In short, the HDE construction has made two assumptions.
(i).\,Holographic principle applies to the entire universe,
so the box size $L$ should be related to the horizon scale in cosmology;
(ii).\,The dark energy is indeed the quantum vacuum energy,
so the dark energy density
$\rho_{\Lambda}^{}\!\sim\! \Lambda^{4}$.
Ref.\,\cite{Li:2004rb} subsequently proposed that,
to make the largest $L$ saturate the above new condition,
the energy density of this HDE should be
\beqa
\rho _{\mathrm{de}}^{}=3c^{2}M_{\mathrm{Pl}}^{2}L^{-2}\,,
\label{eq:rho_hde}
\eeqa
where $\,c\,$ is a constant coefficient.
It was also found\,\cite{Li:2004rb,Ma09} that only if the IR cutoff $L$
is taken as the future event horizon of the Universe,
$\,L\!=\!R_{\rm eh}^{}\!=a\!\int^{\infty}_{t}\!{\rm d}t'/a(t')$\,,\,
the dark energy can provide the desired repulsive force
to explain the cosmic acceleration.

\vspace*{1mm}

\begin{figure}
\begin{center}
\includegraphics[width=0.48\textwidth]{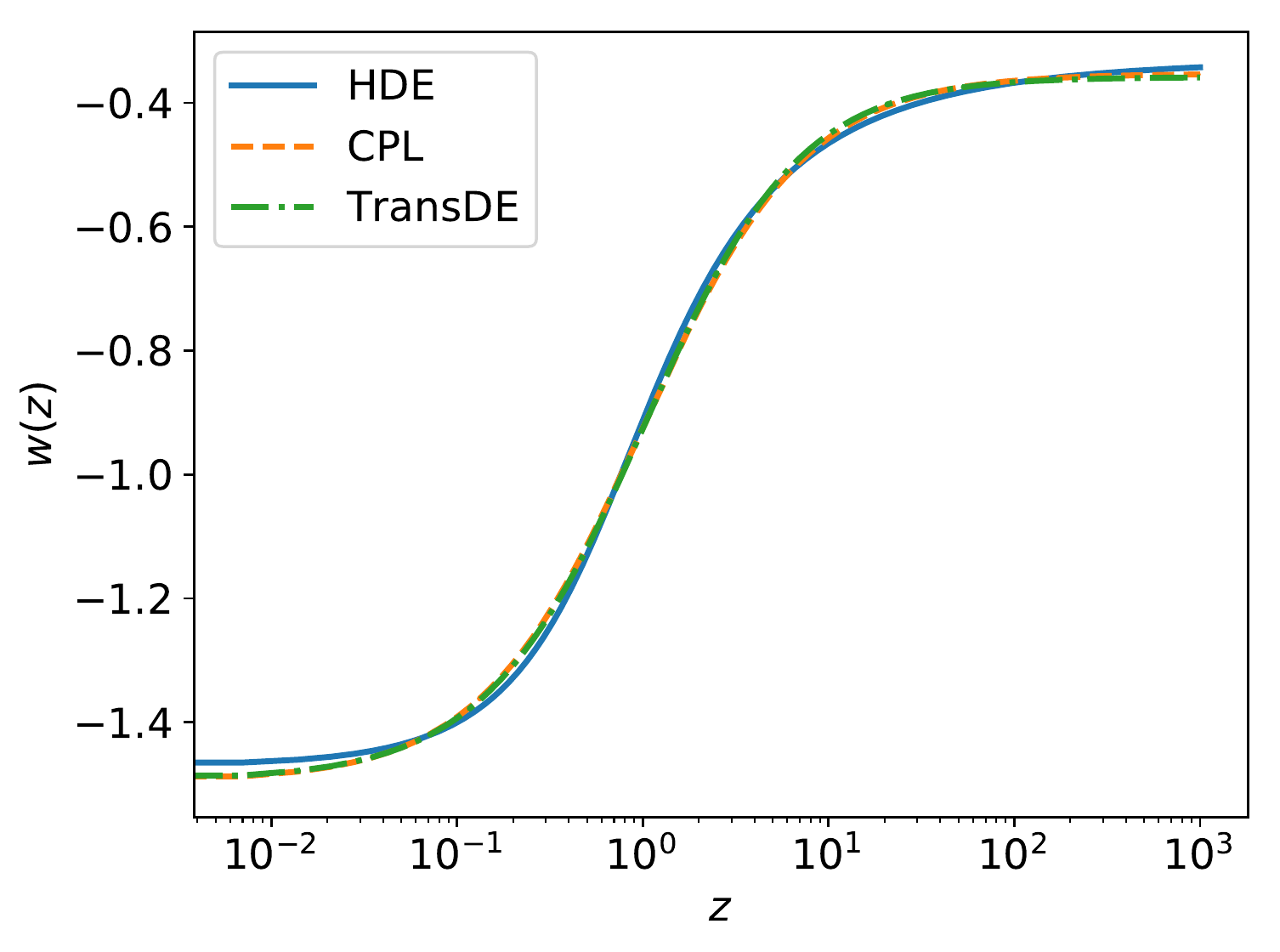}
\vspace*{-8mm}
\caption{Fitting HDE behavior by using TransDE and CPL dark energy models.
The blue solid curve shows the EoS of HDE with $c= 0.5$
and the other physical parameters fixed to the typical {\it Planck} best-fitting values\,\cite{Aghanim:2018eyx}. The orange dashed and green dash-dotted curves are
the best-fitting $w(z)$ of the CPL dark energy and the TransDE, respectively,
in the redshift range $z\in(0, 10^3)$.}
\label{fig:hde_fitting}
\label{fig:1}
\end{center}
\vspace*{-4mm}
\end{figure}

We combine Eq.\eqref{eq:rho_hde} with the energy-momentum conservation
and Friedmann equation to obtain:
\beqs
\label{eq:EMC-FEq}
\begin{eqnarray}
&& \dot{\rho}_{\rm de}^{} + 3 H (1 + w_{\rm de}^{}(z))\rho_{\rm de}^{} =0\,,
\label{eq:cons_de}
\\[1.5mm]
&& 3 M_{\rm Pl}^2 H^2 = \sum_{j}\rho_{j}^{},
\label{eq:Friedmann}
\end{eqnarray}
\eeqs
where $w_{\rm de}^{}(z)$ is the EoS parameter of the HDE
and $H$ is the Hubble parameter.
In Eq.\eqref{eq:Friedmann},
the sum of energy densities includes matter ($\rho_{\rm m}^{}$),
radiation ($\rho_{\rm r}^{}$), and dark energy ($\rho_{\rm de}^{}$).
Among these, $\rho_{\rm r}^{}\!=\Omega_{\rm r}\rho_{\rm cr}(1\!+\!z)^{4}$\,
is fixed by the observed CMB temperature.
From Eq.\eqref{eq:EMC-FEq},  we derive
the following differential equations governing
the dynamics of background expansion,
\beqs
\begin{eqnarray}
\frac{{\di} \rho_{\rm de}^{}}{{\rm d} t}
&=& -2 H \rho_{\rm de}^{}\!
\(\! 1 - \frac{\rho_{\rm de}^{1/2}}{\,\sqrt{3}\,c\,M_{\rm Pl}^{}H\,}\!\)\label{eq:drho-de},
\hspace*{10mm}
\\[1mm]
\frac{{\di} H}{{\di} t} &=&
\frac{1}{\,6 M_{\rm Pl}^2H\,}\sum_j\dot{\rho}_{j}^{} \,,
\label{eq:differential}
\end{eqnarray}
\eeqs
and the EoS for the HDE,
\beqa
w_{\rm de}^{}(z) \,=\,
-\frac{1}{3} - \frac{2}{3}
\frac{\rho_{\rm de}^{1/2}}{\,\sqrt{3}\,c\,M_{\rm Pl}^{}H\,} \,.
\label{eq:EoS-HDE}
\eeqa
We numerically solve Eqs.\,(\ref{eq:drho-de})-(\ref{eq:differential}) as the background evolution of the Universe,
and compute the HDE EoS from Eq.~(\ref{eq:EoS-HDE}). We show the $w_{\rm de}(z)$ function for the case $c=0.5$ as the blue solid curve in Fig.\,\ref{fig:hde_fitting}.

\vspace*{1mm}

In comparison with the ``vanilla'' $\Lambda$CDM cosmology,
the HDE scenario has one extra free parameter $c$ as Eq.(\ref{eq:rho_hde}),
which controls the behaviour of HDE.
Given an appropriate $c$\,, the HDE can have $w_{\rm de}^{}$
greater than $-1$ (corresponding to Einstein's cosmological constant $\Lambda$)
at early epoch ($z \gtrsim 1$), transits to be less than $-1$ at later epoch ($z\lesssim 1$).
This behaviour suggests that, compared to $\Lambda$, the repulsive force in HDE
(quantified as the pressure $P\!=\!w_{\rm de}^{}\rho$\,)
was weaker at the earlier epoch than the present time.
Hence, it causes the Universe to have smaller acceleration earlier on, and faster acceleration at the later stage, but still keeps the total angular diameter distance to the last-scattering surface unchanged.
We find that this ``{\it delayed acceleration}''
is precisely the dynamical behaviour needed
to resolve the $H_0^{}$ tension
because the present-day expansion rate $H_{0}^{}$
from the local measurements
is higher than what is measured by the CMB.
But the angular diameter distance to the last-scattering surface
is fixed by the high-precision CMB measurement.
As an analogue, a marathon runner can run slower at an early stage
but accelerate at the later period to keep the total time and distance unchanged.

\vspace*{4mm}
\noindent
{\bf\large 3.\,Two Parametrized Models}
\vspace*{2.5mm}

To explore the transiting $w(z)$ behaviour,
we seek two parametrized models of the dynamical dark energy
with $2$ and $4$ more parameters than $\Lambda$CDM,
which can mimic the behaviour of HDE.
In general,
if a dark energy model has EoS $\,w_{\rm de}^{}\!>\! -1$\,
at early epoch and transits to $\,w_{\rm de}^{}\!<\! -1$\, at late epoch,
it should have the potential to imitate the HDE.
We first propose a ``TransDE'' parametrization with four free parameters
($w_1^{}, w_2^{}, z_\mathrm{t}^{}, \Delta z$),
\beqa
w(x\equiv \ln(1\!+\!z)) \,=\, w_1^{} \!+\! \frac{w_2^{}}{2}
\!\left(\!1 \!+ \tanh\!\frac{\,x \!-\! x_{\mathrm{t}^{}}\,}{\Delta x\,}\!\right)\!,
~~~~~~~
\label{eq:transDE_w}
\eeqa
where $\,x_\mathrm{t}^{} \equiv \ln ( 1 \!+\! z _\mathrm{t}^{})$\,
and $\,\Delta x \equiv \Delta z /( 1 \!+\! z _\mathrm{t}^{})$\,
determine the redshift $z_{\rm t}^{}$
and width $\Delta z$ of the transition.
The ($w_{1}^{}\!+w_{2}^{}$) and $w_{1}^{}$
control the asymptotic behaviour of EoS at the infinite past
($z \!\to\! \infty$) and infinite future ($z\!\to\! -1$),
respectively.
The other model is the famous Chevallier-Polarski-Linder (CPL) parametrization~\cite{Chevallier:2000qy,PhysRevLett.90.091301},
$\,w(a)= w_0^{} \!+ w_a^{}(1\!-a)\,$,\,
which behaves like the TransDE model at high-$z$,
but the difference is non-negligible if a rapid transition of EoS happens
at low-$z$ \cite{Linden:2008mf}.

\vspace*{1mm}

We substitute the EoS of the TransDE and CPL parametrizations into Eq.(\ref{eq:cons_de})
and obtain the analytical solution for the dark energy density individually,
\beqs
\begin{eqnarray}
\hspace*{-5mm}
&& \rho_\mathrm{de}^\mathrm{TransDE} =\,
\frac{\rho_\mathrm{de}^0}{\,\cosh\!
\left(\! \frac{x_\mathrm{t}}{\Delta x}\!\right)^{\!\!\frac{\,3w_2^{}\Delta x\,}{2}}}
\exp\!\left[3\!\(\!1\!+\!w_1^{} \!+\!\frac{w_2^{}}{2}\!\)\!x\right.
\nonumber \\
&& \hspace*{18mm}
\left.+\frac{3}{2} w_2^{} \Delta x \ln\!
\(\!\cosh\!\frac{\,x \!-\! x_\mathrm{t}^{}\,}{\Delta x}\!\)\right]\!,
\\[1mm]
&&
\rho_\mathrm{de}^\mathrm{CPL} \!=
\rho_\mathrm{de}^0 \exp\!\left\{-3 \LB w_a^{}(1\!-\!a)
\!+\! (1\!+\!w_0^{}\!+\!w_a^{})\ln a\RB\right\}\!,
\hspace*{10mm}
\end{eqnarray}
\eeqs
\hspace*{-2mm}
where $\,\rho_\mathrm{de}^0\!=\Omega_{\rm de}^{}\rho_{\rm cr}^{0}$\,
is the present-day dark energy density, and
$\,\rho_{\rm cr}^{0}\!=3H^{2}_{0}M_{\text{Pl}}^2$\, is the
critical energy density at present.
We fit the HDE scenario with \,$c=0.5$\,
by using TransDE and CPL models respectively, shown in Fig.\,\ref{fig:hde_fitting}. Due to extra free parameters,
both TransDE and CPL models can mimic the HDE behavior,
with the minimal deviations found by the global optimizer
{\tt PyGMO}\,\cite{Pygmo:117}.
This comparison shows that
the HDE model is the most economical model
to resolve the $H_0^{}$ discrepancy.

\vspace*{4mm}
\noindent
{\bf\large  4.\,Data Analysis}
\vspace*{2.5mm}

We combine the R19 data
(local measurement), galaxy baryon acoustic oscillation (BAO) data (median redshifts), and {\it Planck}\, CMB data (high redshifts) in our data fitting.

\begin{figure}[t]
\begin{center}
\includegraphics[width=0.48\textwidth]{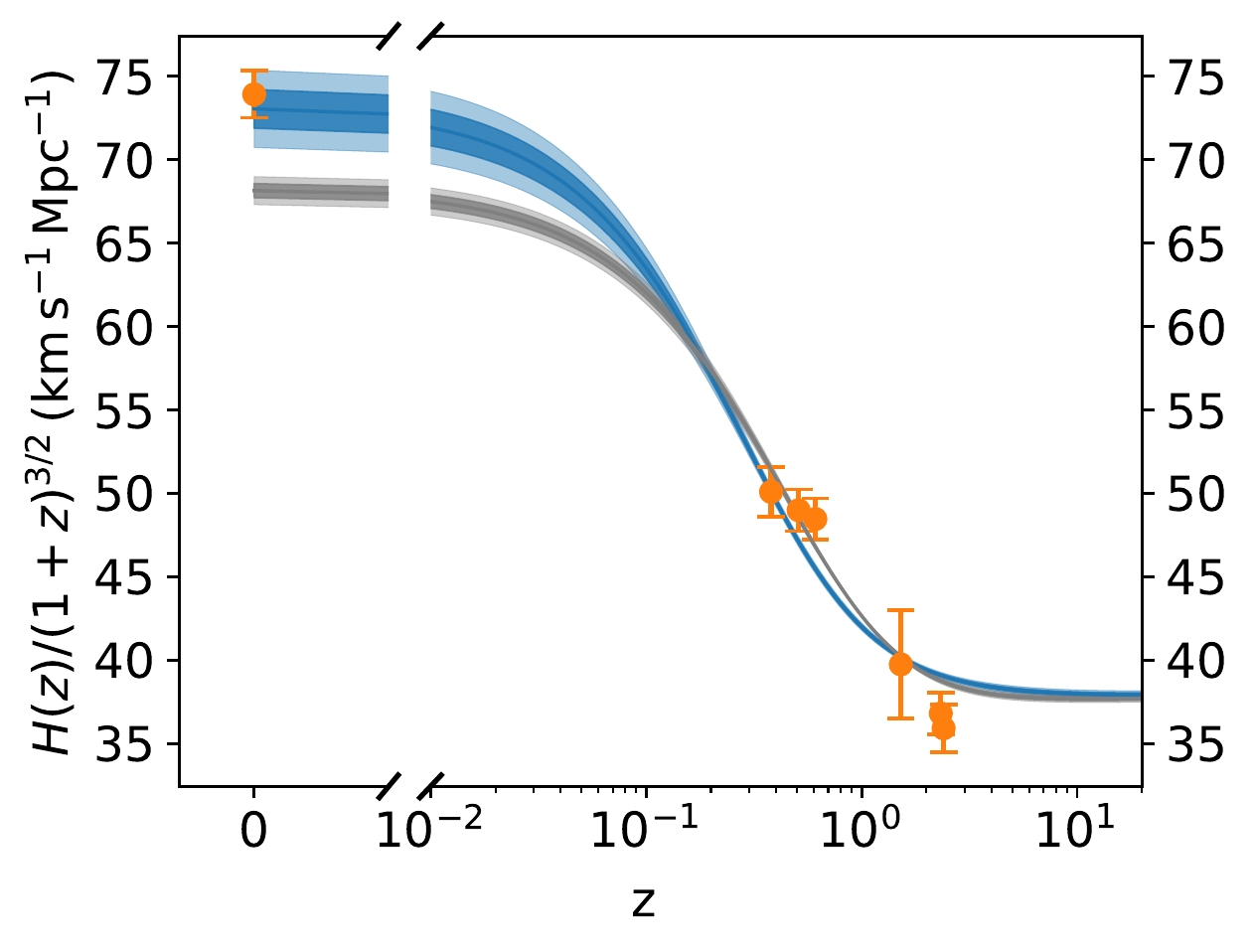}
\vspace*{-8mm}
\caption{{\it Planck}+BAO12+R19 constraints on the Hubble parameter
for HDE (blue) and $\Lambda\mathrm{CDM}$ cosmology (grey).
The dark (light dark) colored stripes present the $68\%$ ($95\%$) limits,
and the black solid curve in the center corresponds to the mean value.
The orange dots with $68\%$ error bars are the (marginalized) measurements.
From left to right, the first point is R19, the next three points are BAO DR12 constraints, and the last three points are the eBOSS DR14 QSO, BOSS DR12 $\mathrm{Ly}\alpha$ and BOSS DR12 $\mathrm{QSOxLy}\alpha$,
which are listed in table\,1 of Ref.\,\cite{Lemos:2018smw}.
The BAO data at $\,z\!\gtrsim\!1\,$ cannot help distinguish
the two models because they are close to each other.}
\label{fig:HubbleParams}
\end{center}
\vspace*{-4mm}
\end{figure}

\vspace*{1mm}

We use the final full-mission baseline {\it Planck} likelihood data (the 2018 release), which includes the low-$\ell$ temperature likelihood (Commander), low-$\ell$ {\it EE} likelihood ({SimAll}), high-$\ell$ {\it TT, TE and EE} likelihood ({Plik})\,\cite{Aghanim:2019ame}, and the additional CMB lensing likelihood\,\cite{Aghanim:2018oex}. In the following, {\it ``Planck''} denotes the combination of the aforementioned {\it Planck} data.

\vspace*{1mm}

The BAO data includes the ``consensus'' SDSS/DR12 data\,\cite{Alam:2016hwk}, the 6dF\,\cite{Beutler:2011hx} data and MGS\,\cite{Ross:2014qpa} BAO data.
Besides, SDSS quasar data and the combination of Lyman-$\alpha$ auto-correlation and Quasar-Lyman-$\alpha$ cross-correlation data have put BAO constraints at redshifts
$\,z\!>\!2$ \cite{Ata:2017dya,Agathe:2019vsu,Blomqvist:2019rah} which we plotted in Fig.\,\ref{fig:HubbleParams} ($z \!\gtrsim\! 2$) for comparison. But unlike galaxy BAO measurements, quasar Ly$\alpha$ measurements require several additional assumptions of the modeling of metal-line and high-column-density neutral hydrogen and quasar spectra universality, hence are more complicated than the galaxy BAO measurements.
Besides, the HDE and $\Lambda$CDM are nearly indistinguishable
for the Hubble parameter in the corresponding redshift range.
For these reasons, we do not include the Ly$\alpha$ BAO in the parameter constraints,
but use 6dF, MGS and SDSS/DR12 data as ``BAO12''. (Ref.~\cite{Aghanim:2018eyx} gives similar strategy and reason)

\begin{figure}[t]
\begin{center}
\includegraphics[width=0.48\textwidth]{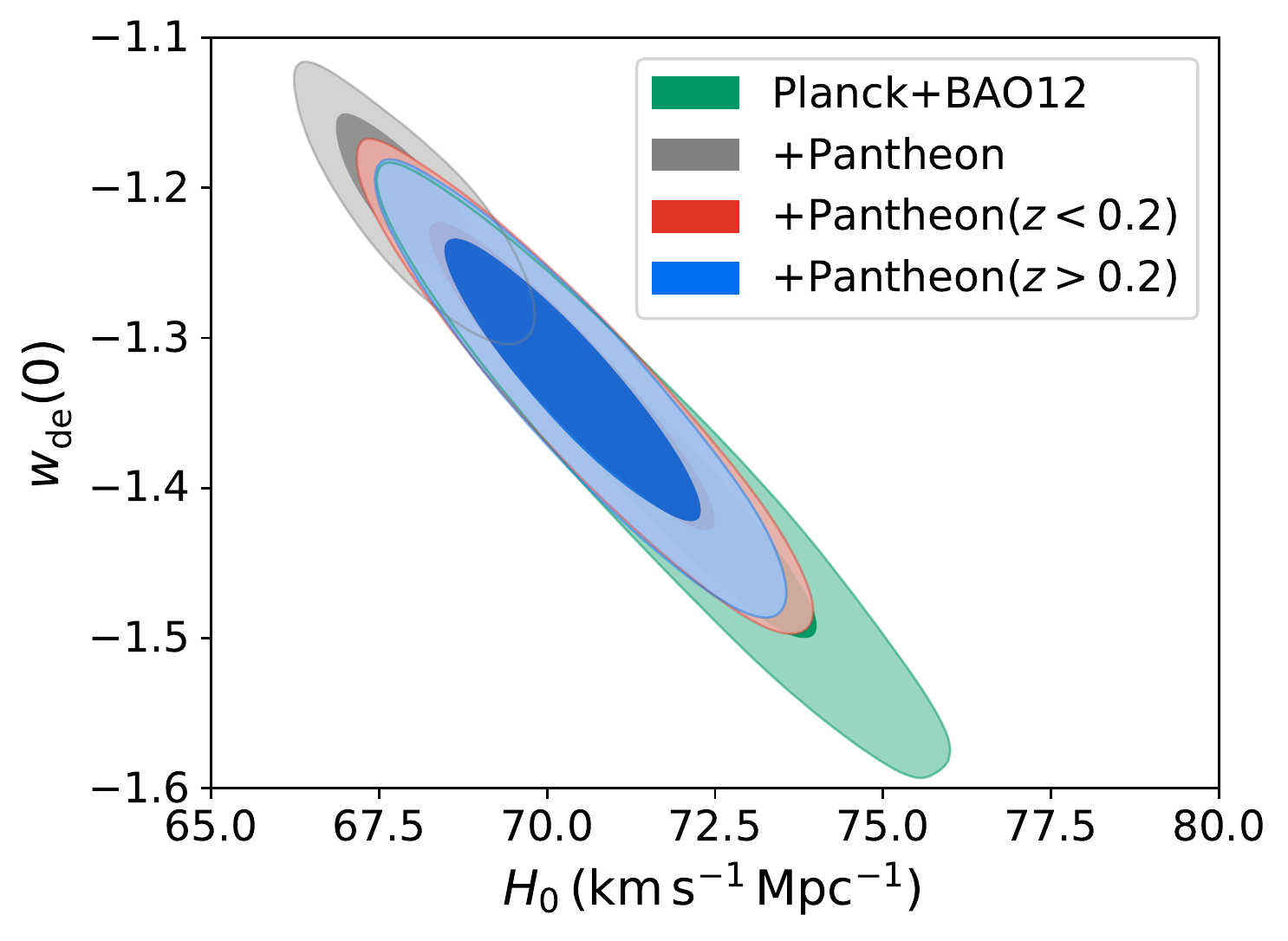}
\vspace*{-8mm}
\caption{Joint constraints for the HDE on $w_{\rm de}(0)$
(projected EoS at $z=0$) and $H_{0}$ from
{\it Planck} + BAO12, {\it Planck} + BAO12+Pantheon ($z\!<\!0.2$),
{\it Planck} + BAO12 + Pantheon ($z\!>\!0.2$),
and {\it Planck} + BAO12 + Pantheon (full).
It shows that {\it Planck} + BAO12 + partial Pantheon dataset
(with either $z>0.2$ or $z<0.2$) gives a higher value of $H_0^{}$,
but with the full dataset the value becomes much lower.
This is because the Pantheon dataset has
{large correlation} between high-$z$ and low-$z$ samples
(cf.\ text).}
\label{fig:pantheon-w-H0}
\end{center}
\vspace*{-6mm}
\end{figure}

\vspace*{1mm}

Table\,\ref{tab:bao} enumerates the effective redshift for each measurement,
ranging from $0.106$ to $0.61$.
The parameter $r_{\mathrm{d}}^{}$ is the sound horizon at drag epoch
and $D_{\mathrm{M}}^{}$ denotes the comoving angular diameter distance.
$D_{\mathrm{V}}^{}$ is determined by the angular diameter distance
$D_{\rm A}^{}$ and the Hubble parameter $H(z)$ via
$\,D_{\mathrm{V}}^{}\!=\![c D_{\rm A}^2 z(1\!+\!z)^2\!/H(z)]^{1/3}$.\,
The 6dF and MGS data give the measurement of
$\,r_\mathrm{d}^{}/D_\mathrm{V}^{}$\,
at redshift $\,z_\mathrm{eff}^{}\!=\!0.106$\,
and the measurement of
$\,D_\mathrm{V}^{}/r_\mathrm{d}^{}$
at redshift $z_\mathrm{eff}^{}\!=\!0.15$,\, respectively.
BOSS DR12 data include
$D_{\mathrm{M}}^{}r_\mathrm{fid, d}^{}/r_\mathrm{d}^{}$
and $H r_\mathrm{d}^{}/r_\mathrm{fid, d}^{}$
at redshifts $\,z_\mathrm{eff}^{}\!=\!\{0.38, 0.51$, $0.61\}$,\,
where $\,r_\mathrm{fid, d}^{} \!=\! 147.78\,\mathrm{Mpc}$\,
is a fiducial sound horizon.
Since DR12 data are correlated between different redshifts,
we include all their full covariance matrix in our {\sc CosmoMC} likelihood package.

\vspace*{1mm}

R19 is the measurement of $H_{0}$ from Large Magellanic Cloud Cepheid Standards
by Riess {\it et al.}\,\cite{Riess:2019cxk},
which gives $H_0^{} \!=\! 74.03\pm1.42\,\mathrm{km \,s^{-1} Mpc^{-1}}$,
deviating from {\it Planck} measurement at $4.4\sigma$ level.
Pantheon is a new set of light-curve supernovae (SNe),
with $1048$ samples spanning the redshift range $\,0.01\!<\!z\!<\!2.3$ \cite{Scolnic:2017caz}.
The SN samples do not directly measure $H_{0}$
because it is degenerate with the absolute magnitude $M$.
The constraint comes indirectly from the joint datasets with
{\it Planck}, because the SN samples
can put constraints on $\Omega_{\rm m}$
(fractional matter density) and $w_{\rm de}$
that have covariance with $H_0^{}$\,.
In Fig.\,\ref{fig:pantheon-w-H0}, we plot the joint constraint of
$w_{\rm de}(0)$
(projected dark energy EoS at $z\!=\!0$\,)
and $H_0^{}$ from {\it Planck}+BAO12,
{\it Planck} $\!+\!$ BAO12 $\!+\!$ Pantheon ($z\!<\!0.2$ subset),
{\it Planck} $\!+\!$ BAO12 $\!+\!$ Pantheon ($z\!>\!0.2$ subset),
and {\it Planck} $\!+\!$ BAO12 $\!+\!$ Pantheon (full dataset).
It shows that the combined {\it Planck}+BAO12
with either subset of the Pantheon data gives consistent results
of higher $H_0^{}$ value, but the full Pantheon samples shift to a lower value.
We find that this inconsistency between the full Pantheon samples
and each subset of samples is due to the {\it large correlation}
between high-$z$ and low-$z$ samples of Pantheon.
In Ref.\,\cite{DiValentino20}, it shows that {\it Planck}+BAO and
{\it Planck}+Pantheon give inconsistent results at more than 95\% C.L.,
suggesting that there are uncounted
systematics in either Pantheon or R19 data.
Since there are several local measurements that support R19 results
(e.g., TRGB and lensing), we will not adopt {\it Planck}+BAO+Pantheon
as a baseline dataset, instead we use {\it Planck} $\!+\!$ BAO12 and
{\it Planck} $\!+\!$ BAO12 $\!+\!$ R19 as two baseline datasets.

\begin{figure}[t]
\begin{center}
\includegraphics[width=0.48\textwidth]{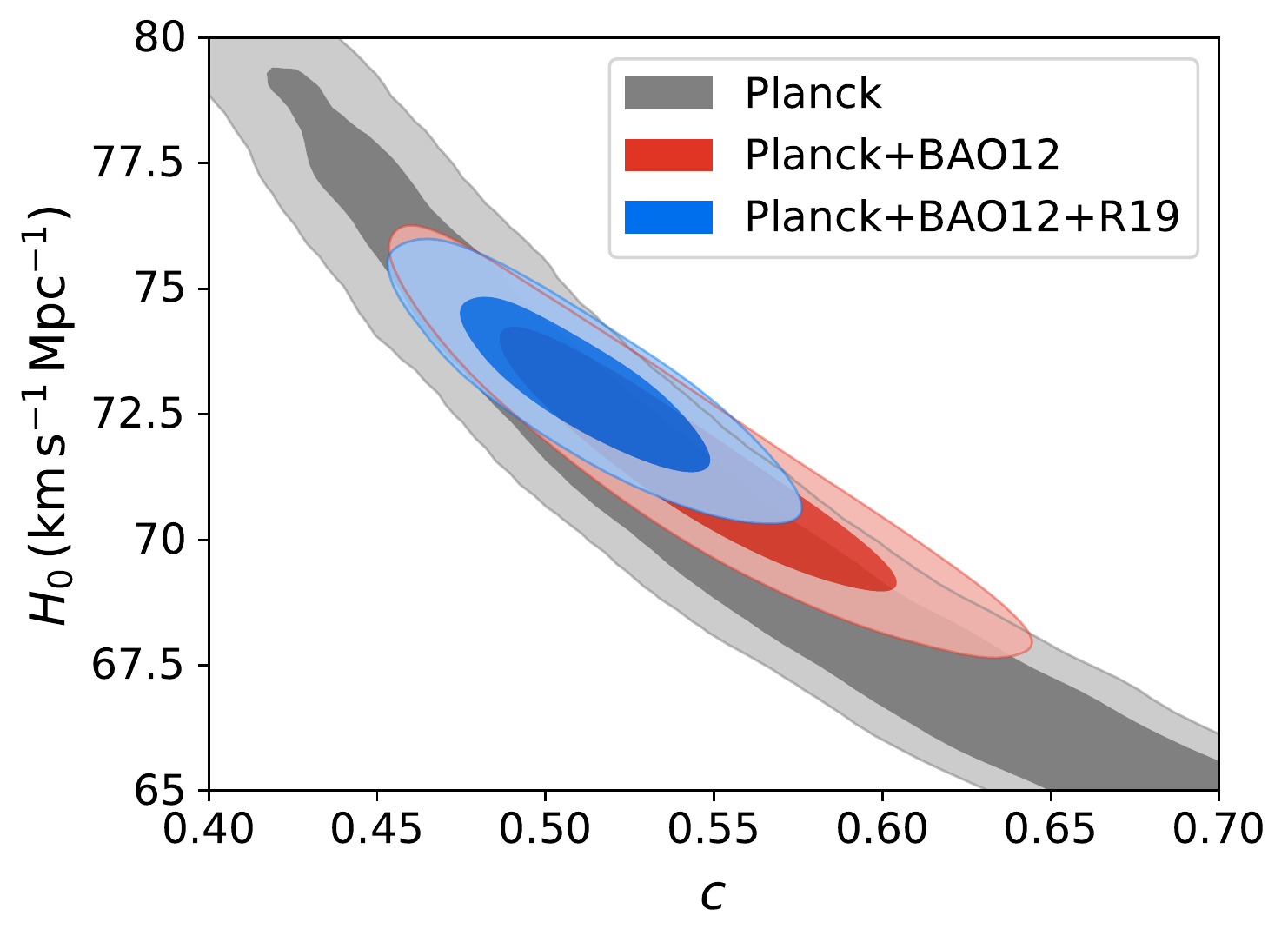}
\vspace*{-9mm}
\caption{Marginalized constraints on the Hubble constant $H_{0}$ versus
the HDE parameter $c\,$, at $68\%$\,C.L.\,(contours with dark colors)
and $95\%$\,C.L.~(contours with light colors).
The combinations of three datasets are shown in the legend.}
\label{fig:c_hde}
\end{center}
\vspace*{-4mm}
\end{figure}

\vspace*{4mm}
\noindent
{\bf\large 5.\,Results and Discussions}
\vspace*{1.5mm}

We modify the Boltzmann
{\sc camb} code\,\cite{Lewis:1999bs}
to embed the HDE and TransDE models into the background expansion
of the Universe, and use public code {\sc CosmoMC} (version of July\,2019)
to explore the parameter space with Markov chains Monte Carlo
technique\,\cite{Lewis:2002ah}. We assume spatially flat cosmology with six base cosmological parameters
($\Omega_{\rm c}^{}h^{2}$, $\Omega_{\rm b}^{}h^{2}$, $\theta_{\ast}^{}$,
$n_{\rm s}^{}$, $A_{\rm s}^{}$, $\tau$) to vary,
and set only one of the three generations of neutrinos having mass $0.06\,$eV
(under normal neutrino mass ordering).
We then add the dark energy sector into the analysis which has one,
two and four dark energy parameters for the HDE, CPL and TransDE models,
respectively.

\vspace*{1mm}

Figure\,\ref{fig:c_hde} presents the marginalized 2D contour of
the HDE parameter $\,c\,$ versus $H_0^{}$\,.
The {\it Planck}-only constraint on $H_0^{}$ is relatively weak,
but including the BAO12 and BAO12+R19 data tightens up the bounds.
In Fig.\,\ref{fig:HubbleParams}, we plot the evolution of Hubble parameter
$H(z)$ as a function of the redshift within range $\,z\!\in\![0,20]$\,
for the HDE (blue) and $\Lambda$CDM cosmology (grey)
under $1\sigma$ and $2\sigma$ variations.
We see that the ``delayed acceleration'' effect of HDE can match
the BAO data and local R19 data better than $\Lambda$CDM model.
This fact is also reflected by the $\chi^{2}_{\rm min}$ and $\Delta$AIC values
listed in Table\,\ref{tab:aic}.
The Akaike information criterion (AIC) is a metric to quantify the
``goodness-of-fit'' by compensating the additional parameter(s) in the model.
Comparing the HDE (1 extra parameter), TransDE (4 extra parameters) and CPL
(2 extra parameters) with the benchmark $\Lambda$CDM cosmology, the
HDE model fits the data better than the TransDE and $\Lambda$CDM models,
while it also predicts the $H_0^{}$ value consistent with both the local R19
and strong lensing measurements.
Although the CPL model performs better in terms of the AIC,
it does not recover the local $H_0^{}$ value (Table~\ref{tab:aic}).

\begin{table}[t] 
\renewcommand\arraystretch{1.2}
\begin{center}
\caption{BAO measurements.
$D_\mathrm{V}^{}$, $D_\mathrm{M}^{}$ $D_\mathrm{H}^{}$ and Hubble parameter $H$
are computed at the effective redshifts $z_\mathrm{eff}^{}$\,.}
\vspace*{1.5mm}
\dtabsize
\begin{tabularx}{0.48\textwidth}{p{0.07\textwidth}p{0.05\textwidth}p{0.12\textwidth}X}
\hline
\hline
Dataset &$z_{\mathrm{eff}}$ &Measurement & Constraint \\
\hline
6dF & $0.106$   &$r_{\mathrm{d}}/D_{\mathrm{v}}$ &$0.336\pm0.015$ \\
MGS & $0.15$   &$D_{\mathrm{v}}/r_{\mathrm{d}}$ &$4.47\pm0.17$\\
SDSS/ & $0.38$ &$D_{\mathrm{M}}r_{\mathrm{fid, d}}/r_{\mathrm{d}}$ &$ (1512.39\pm24.99) ~\mathrm{Mpc}$\\
DR12     &       &$H r_{\mathrm{d}}/r_{\mathrm{fid, d}}$ &$(81.21\pm2.37) ~\mathrm{km ~s^{-1} Mpc^{-1}}$ \\
     &$0.51$ &$D_{\mathrm{M}} r_{\mathrm{fid, d}}/r_{\mathrm{d}}$ &$(1975.22\pm30.10) ~\mathrm{Mpc}$\\
     &       &$H r_{\mathrm{d}}/r_{\mathrm{fid, d}}$ &$(90.90\pm2.33) ~\mathrm{km ~s^{-1} Mpc^{-1}}$ \\

     &$0.61$ &$D_{\mathrm{M}} r_{\mathrm{fid, d}}/r_{\mathrm{d}}$ &$(2306.68\pm37.08) ~\mathrm{Mpc}$\\
     &       &$H r_{\mathrm{d}}/r_{\mathrm{fid, d}}$ &$(98.96\pm2.50) ~\mathrm{km ~s^{-1} Mpc^{-1}}$ \\
\hline\hline
\end{tabularx}
\label{tab:bao}
\end{center}
\vspace*{-4mm}
\end{table}

\begin{table}[t]
\vspace*{-2mm}
\renewcommand\arraystretch{1.2}
\begin{center}
\caption{Model comparison.
$\Delta\mathrm{AIC}$ is the difference of $\mathrm{AIC}$ from
the $\Lambda\mathrm{CDM}$ model with the same dataset.}
\vspace*{1.5mm}
\dtabsize
\begin{tabularx}{0.48\textwidth}{XXccc}
\hline
\hline
\multirow{2}{*}{Data\,Set} &\multirow{2}{*}{Model}
& Best-fitting $H_0^{}$ &\multirow{2}{*}{~~~$\chi^2_{\min}$~~~}
&\multirow{2}{*}{~~$\Delta\rm{AIC}$~~} \\
& &{\footnotesize{$[\mathrm{km}\,\mathrm{s}^{-1}\mathrm{Mpc}^{-1}]$}} \\
\hline
\multirow{4}{*}{\shortstack[l]{{\it Planck}\\+BAO12\\+R19}} &HDE     &$73.47$ &$2791.58$ &$-5.67$\\
&TransDE &$71.40$ &$2789.38$ &$-1.87$\\
&CPL     &$71.60$ &$2787.75$ &$-7.5$\\
&$\Lambda\mathrm{CDM}$ &$68.23$ &$2799.25$ &$0$\\
%
\hline\hline
\end{tabularx}
\label{tab:aic}
\end{center}
\vspace*{-4mm}
\end{table}

\vspace*{1mm}

\begin{figure}[t]
\begin{center}
\includegraphics[width=0.48\textwidth]{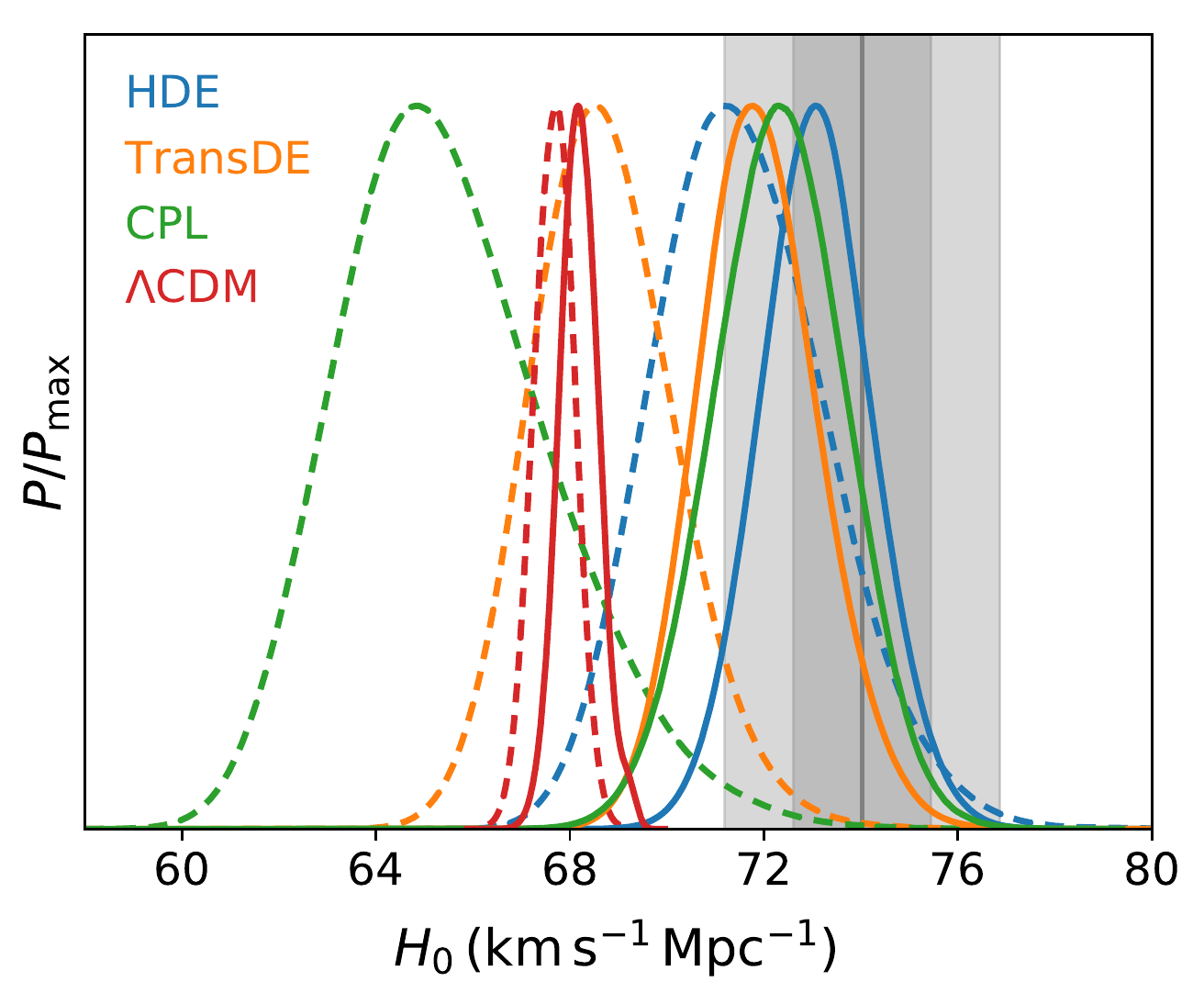}
\vspace*{-8mm}
\caption{Marginalized distributions of the Hubble parameter $H_0^{}$.
The solid (dashed) curves show the {\it Planck}+BAO12+R19
({\it Planck}+BAO12) constraints.\
The vertical bands denote allowed regions by the R19 measurements at
$1\sigma$ (dark grey) and $2\sigma$ (light grey) confidence levels.}
\label{fig:h0_dist}
\end{center}
\vspace*{-4mm}
\end{figure}

Figure\,\ref{fig:h0_dist} shows the projected distribution of $H_0^{}$
in the HDE, TransDE, CPL, and $\Lambda\mathrm{CDM}$ cosmologies for both {\it Planck}+BAO12 (dashed curves) and
{\it Planck}+BAO12+R19 (solid curves) datasets.
One can see that with R19 data (solid), all three
dark energy models prefer higher values of $H_0$
which are consistent with the vertical grey bands (R19 data).
But without R19 data (dashed), the TransDE and CPL dark
energy models shift its projected $H_0$ value to lower values and become less consistent with the R19 value. Hence, it is the R19 data that dictates the TransDE and CPL models to have higher values of $H_0$, without which the recovery does not exist. However, even without the R19 data, the HDE predicts a projected value of $H_0=
71.54 \!\pm\! 1.78\,\mathrm{km \,s^{-1} Mpc^{-1}}$, which is fully consistent with the R19 data within the $1.4\sigma$ range.
This resolution is due to its inherent behaviour of $w(z)$.
With {\it Planck}+BAO12+R19 results (solid curves) in Fig.\,\ref{fig:h0_dist},
we derive Hubble constant
$\,H_0^{} \!= 73.12\!\pm\!1.14 \,\mathrm{km \,s^{-1} Mpc^{-1}}$ and the input parameter
$\,c = 0.51\!\pm\!0.02$.\,
Hence, the combined constraints for HDE give the closest value of $H_0^{}$
to the R19 and strong lensing measurements. This fully resolves the
$H_0^{}$ tension between the CMB and local measurements.

\vspace*{1mm}

Finally, we emphasize that, in contrast to the CPL and TransDE parametrizations
and other phenomenological approaches,
the HDE is physically well motivated from the
holographic principle\,\cite{tHooft:1993dmi}-\cite{hawking1975}
of quantum gravity that connects the total energy of vacuum state
to the cosmic horizon scale (as the infrared cutoff)\,\cite{Cohen99}\cite{Li:2004rb}.\
It naturally provides dynamical dark energy with only one free parameter that keeps the total angular diameter distance to LSS unchanged, while it increases the local expansion. This behaviour resolves the $H_0^{}$ discrepancy
between the CMB and local $H_0^{}$ measurements
in an exquisite and economical way.
More importantly, the HDE model is generically different from
the $\Lambda$CDM due to its transiting equation of the state.
Thus future measurements will improve the constraints and possibly substantiate
or falsify the HDE from the benchmark $\Lambda$CDM Universe.
This is evident in Fig.\,\ref{fig:HubbleParams} that the major difference between the HDE and $\Lambda$CDM lies
at the redshift range $\,0\!<\! z \!<\!1$\,. The DESI survey\,\cite{DESI-collaboration}, which aims to collect 36 million galaxy spectra by 2026, will undoubtedly put strong constraints
on the cosmic evolution at this epoch.
Besides, new probes such as the gravitational wave\,\cite{Hotokezaka19}
and BAO measurement from 21-cm intensity mapping\,\cite{Zhang19}
will further improve the constraints significantly.

\vspace*{4mm}
\noindent
{\bf Acknowledgements}
\\[1mm]
Y.Z.M. acknowledges the support of NRF-120385, NRF-120378, and NSFC-11828301.
H.J.H. was supported by NSF of China
(No.\,11675086 and No.\,11835005), CAS Center for Excellence
in Particle Physics (CCEPP), National Key R\,\&\,D Program of
China (No.\,2017YFA0402204), by the Key Laboratory for Particle Physics,
Astrophysics and Cosmology (MOE),
and by the Office of Science and Technology, Shanghai Municipal Government
(No.\ 16DZ2260200).


\bibliography{hde_ref}

\end{document}